\documentclass[aps,reprint,superscriptaddress]{revtex4-1} 
\usepackage[english]{babel}
\usepackage{amsmath, amssymb}
\usepackage{mathtools}
\usepackage{siunitx}
\usepackage{environ}
\usepackage{mathrsfs}
\usepackage{braket}
\usepackage{bm}
\usepackage{prettyref}
\usepackage{graphicx}
\usepackage{booktabs}
\usepackage{array}
\usepackage{multirow}
\usepackage{dcolumn}
\usepackage{threeparttable}
\usepackage{threeparttablex}
\usepackage{placeins}
\usepackage{pgf}
\usepackage{tikz}

\newcommand{\pauli}{\boldsymbol{\sigma}}
\newcommand{\Pauli}{\boldsymbol{\sigma}}
\newcommand{\diraccontra}[1]{\boldsymbol{\gamma}^{#1}}

\newcommand{\pos}{\vec{r}}

\newcommand{\momop}{\hat{\vec{p}}}

\newcommand{\spinmom}{\vec{\Pauli}\cdot\momop}

\newcommand{\Sum}[2]{\sum\limits_{#1}^{#2}}
\newcommand{\parantheses}[1]{\left(#1\right)}
\newcommand{\brackets}[1]{\left[#1\right]}
\newcommand{\braces}[1]{\left\{ #1\right\}}
\let\nablatmp\nabla
\renewcommand{\nabla}{\vec{\nablatmp}}
\DeclarePairedDelimiter\abs{\lvert}{\rvert}
\makeatletter
\let\oldabs\abs
\def\abs{\@ifstar{\oldabs}{\oldabs*}}

\newcommand{\Op}[1]{\hat{#1}}

\AtBeginDocument{
\heavyrulewidth=.08em
\lightrulewidth=.05em
\cmidrulewidth=.03em
\belowrulesep=.65ex
\belowbottomsep=0pt
\aboverulesep=.4ex
\abovetopsep=0pt
\cmidrulesep=\doublerulesep
\cmidrulekern=.5em
\defaultaddspace=.5em
}
\begin{document}
\title{\textit{Ab initio} study of parity and time-reversal violation in 
laser-coolable triatomic molecules}
\date{\today}
\author{Konstantin Gaul}
\affiliation{Fachbereich Chemie, Philipps-Universit\"{a}t Marburg, Hans-Meerwein-Stra\ss{}e 4, 35032 Marburg}
\author{Robert Berger}
\affiliation{Fachbereich Chemie, Philipps-Universit\"{a}t Marburg, Hans-Meerwein-Stra\ss{}e 4, 35032 Marburg}
\begin{abstract}
Electronic structure enhancement factors of simultaneous parity and
time-reversal violation ($\mathcal{P,T}$-violation) caused by an
electric dipole moment of the electron (eEDM) and scalar-pseudoscalar
nucleon-electron current (SPNEC) interactions are reported for various
metal mono-hydroxides, several of which are considered laser--coolable
and promising candidates for an eEDM measurement. Electronic
structure enhancements are calculated \textit{ab initio} within zeroth
order regular approximation (ZORA) for CaOH, SrOH, BaOH, RaOH and
YbOH. Scaling behavior with respect to nuclear charge numbers and the
ratio of enhancement factors for both discussed sources of
$\mathcal{P,T}$-violation are analyzed, which are crucial to obtain
stringent bounds on parameters for new physics from experiments. 
\end{abstract}

\maketitle

High precision spectroscopy of diatomic molecules serves as powerful
tool to probe high-energy scales of new physics beyond the standard
model of particle physics.\cite{demille:2015} Signatures of new
physics are expected for instance from simultaneous parity
$\mathcal{P}$ and time-reversal $\mathcal{T}$
violation.\cite{khriplovich:1997} Such a violation of fundamental
symmetries can in principle result in a permanent electric dipole
moment of a molecule in a vanishing electric field. With cold polar
heavy molecules such as ThO currently the strictest limits on
$\mathcal{P,T}$-violating effects are
set.\cite{baron:2014,andreev:2018} This is due to electronic structure
effects in polar heavy diatomic molecules, which strongly enhance
$\mathcal{P,T}$-odd effects such as an electric dipole moment of an
electron (eEDM) $d_\text{e}$ or scalar-pseudoscalar nucleon-electron
current (SPNEC) interactions.\cite{ginges:2004}
Refs.~\cite{isaev:2013,kudashov:2014,sasmal:2016a} highlighted the
particular situation of $\mathcal{P,T}$-odd effects in the diatomic
system RaF, which was earlier identified to have the advantage of
being also a molecular candidate for laser-cooling.\cite{isaev:2010}
Based on simple theoretical concepts \cite{isaev:2016} it was
subsequently concluded that not only diatomic, but also polyatomic
molecules can be cooled with lasers. This renders such molecules
promising laboratories for the study of fundamental symmetry
violations. A number of molecular candidates was
proposed\cite{isaev:2016} which included the particular example of
CaOH. The first successful experiment of laser-cooling of a polyatomic
molecule was subsequently realised with SrOH\cite{kozyryev:2017}.
Isaev et al.~\cite{isaev:2017} suggested laser-cooling of RaOH and its
use to search for new physics. They presented also the first
calculation of SPNEC interactions enhancement in a polyatomic
molecule, but as of yet, no predictions on eEDM enhancement in the
mentioned polyatomic candidates for a search of new physics exist.

Kozyryev et al. elucidated that laser-coolable polyatomic molecules,
and in particular YbOH, can have advantages over diatomic molecules in
experimental setups and may improve sensitivity of eEDM
experiments.~\cite{kozyryev:2017a} And it was pointed out in
Ref.~\cite{gaul:2018} that diatomic molecules are limited in the
sensitivity of a simultaneous determination of $d_\text{e}$ as well as
the coupling constant of SPNEC interactions $k_\text{s}$ when one
analyses the scaling behaviour of the enhancement factors with respect
to the charge of the heavy nucleus.

To provide these enhancement factors for upcoming experiments on
triatomic molecules, we present in this letter predictions of
$W_\text{d}$ and $W_\text{s}$, the electronic structure enhancement
factors of an eEDM and SPNEC interactions, respectively, in the
laser-coolable polyatomic molecules CaOH, SrOH, RaOH and YbOH, as well
as for BaOH, which is isoelectronic to BaF, a promising candidate for
the first detection of molecular parity-violation.\cite{altundas:2018}
We compare herein also the ratio $W_\text{d}/W_\text{s}$ to those
obtained for diatomic molecules in order to gauge the advantage of
polyatomic over diatomic molecules with respect to electronic
structure enhancement effects.


The metal hydroxides (MOH) studied herein are linear molecules and
expected to have a $^2\Sigma_{1/2}$-ground state. Thus the effective
$\mathcal{P,T}$-odd spin-rotational Hamiltonian is the same as for
diatomic molecules studied in Ref. \cite{gaul:2018}, if one neglects
contributions of the light nuclei, namely: \begin{equation}
H_\text{sr}=\parantheses{k_\text{s}W_\text{s}+d_\text{e}W_\text{d}}\Omega,
\label{eq: spinrot} \end{equation} where
$\Omega=\vec{J}_\text{e}\cdot\vec{\lambda}$ is the projection of the
reduced total electronic angular momentum $\vec{J}_\text{e}$ on the
molecular axis, defined by the unit vector $\vec{\lambda}$ pointing
from the heavy nucleus to the OH-group. $k_\text{s}$ is the
$\mathcal{P,T}$-odd scalar-pseudoscalar nucleon-electron current
interaction constant and $d_\text{e}$ is the eEDM. The
$\mathcal{P,T}$-odd electronic structure parameters are defined by
\begin{equation}
W_\text{s}=\frac{\Braket{\Psi|\Op{H}_\text{s}|\Psi}}{k_\text{s}\Omega}~\text{and}~
W_\text{d}=\frac{\Braket{\Psi|\Op{H}_\text{d}|\Psi}}{d_\text{e}\Omega},
\end{equation} where $\Psi$ is the electronic wave function. The
molecular $\mathcal{P,T}$-odd
Hamiltonians\cite{salpeter:1958,martensson-pendrill:1987,khriplovich:1997}
\begin{align} \Op{H}_\text{s}&=\i
k_{\text{s}}\frac{G_\text{F}}{\sqrt{2}}\Sum{i=1}{N_\text{elec}}
\Sum{A=1}{N_\text{nuc}}\rho_A\parantheses{\pos_i}Z_A\diraccontra{0}\diraccontra{5},
\label{eq: eNcpviol}\\ \Op{H}_{\text{d}}&=\frac{2\i c
d_\text{e}}{\hbar e}\Sum{i=1}{N_\text{elec}}
\diraccontra{0}\diraccontra{5}\momop_i^2.  \label{eq: stratagemII}
\end{align} were implemented and evaluated in a quasi-relativistic
framework within zeroth order regular approximation
(ZORA):\cite{gaul:2017,gaul:2018} \begin{align}
\Op{H}^\text{ZORA}_{\text{s}}=&\i\Sum{i=1}{N_\text{elec}}
\Sum{A=1}{N_\text{nuc}}Z_A\brackets{\rho_A(\pos_i)
\tilde{\omega}_\text{s}(\pos_i),\spinmom_i}_- \label{eq: zoraeNPT},\\
\Op{H}^\text{ZORA}_{\text{d}}=&\i\Sum{i=1}{N_\text{elec}}\momop_i^2\tilde{\omega}_{\text{d}}(\pos_i)\parantheses{\spinmom_i}-\parantheses{\spinmom_i}\tilde{\omega}_{\text{d}}(\pos_i)\momop_i^2
\label{eq: ZORA}.  \end{align} Here $\rho_A$ is the normalized nuclear
charge density distribution of nucleus $A$ with charge number $Z_A$,
$\pos_i$ is the position vector of electron $i$,
$G_\text{F}=2.22249\times10^{-14}~E_\text{h}a_0^3$ is Fermi's weak
coupling constant, $\i=\sqrt{-1}$ is the imaginary unit, $\momop$ is
the linear momentum operator, $\vec{\pauli}$ is the vector of the
Pauli spin matrices, $\brackets{A,B}_-=AB-BA$ is the commutator and
the modified ZORA factors are defined as \begin{align}
\tilde{\omega}_\text{s}(\pos_i)&=\frac{G_\text{F}k_\text{s}c}{\sqrt{2}\parantheses{2m_\text{e}c^2-\tilde{V}(\pos_i)}}~~,\\
\tilde{\omega}_{\text{d}}(\pos_i)&=\frac{2d_\text{e}c^2}{2e\hbar
m_\text{e}c^2-e\hbar\tilde{V}(\pos_i)}.  \end{align} with the model
potential $\tilde{V}$ introduced by van W\"ullen\cite{wullen:1998},
which is used to alleviate the gauge dependence of ZORA. Here $c$ is
the speed of light in vacuum, $\hbar=\frac{h}{2\pi}$ is the reduced
Planck constant and $m_\text{e}$ is the mass of the electron.

On the technical side, the quasi-relativistic two-component
calculations reported herein are performed within ZORA at the level of
complex generalized Hartree--Fock (cGHF) or Kohn--Sham (cGKS) with a
modified
version\cite{berger:2005,berger:2005a,nahrwold:09,wullen:2010} of the
quantum chemistry program package Turbomole\cite{ahlrichs:1989}.  For
details on our implementation of $\mathcal{P,T}$-odd properties within
this ZORA framework see Refs. \cite{isaev:2012,gaul:2017,gaul:2018}.
For Kohn--Sham (KS)-density functional theory (DFT) calculations the
hybrid Becke three parameter exchange functional and Lee, Yang and
Parr correlation functional
(B3LYP)\cite{stephens:1994,vosko:1980,becke:1988,lee:1988} was
employed.  For all calculations a basis set of 37~s, 34~p, 14~d and
9~f uncontracted Gaussian functions with the exponential coefficients
$\alpha_i$ composed as an even-tempered series as $\alpha_i=a\cdot
b^{N-i};~ i=1,\dots,N$, with $b=2$ for s- and p-function and with
$b=(5/2)^{1/25}\times10^{2/5}\approx 2.6$ for d- and f-functions was
used for the electro-positive atom (for details see Supplementary
Material).  The N, F and O atoms were represented with a decontracted
atomic natural orbital (ANO) basis set of triple-$\zeta$
quality\cite{roos:2004} and for H the s,p-subset of a decontracted
correlation-consistent basis of quadruple-$\zeta$
quality\cite{dunning:1989} was used.  The ZORA-model potential
$\tilde{V}(\pos)$ was employed with additional damping\cite{liu:2002}
as proposed by van W\"ullen\cite{wullen:1998}. For two-component wave
functions and properties a finite nucleus was used, described by a
normalized spherical Gaussian nuclear density distribution
$\rho_A(\pos)=\rho_0\mathrm{e}^{-\frac{3}{2\zeta_A}\pos^2}$.  The root
mean square radius $\zeta_A$ of nucleus $A$ was used as suggested by
Visscher and Dyall,\cite{visscher:1997} where the mass numbers $A$ are
$^{40}$Ca, $^{137}$Ba, $^{173}$Yb, $^{226}$Ra.  The nuclear
equilibrium distances were obtained at the levels of GHF-ZORA and
GKS-ZORA/B3LYP, respectively. For calculations of energy gradients at
the DFT level the nucleus was approximated as a point charge. The
molecular structure parameters obtained are summarized in
\prettyref{tab: molpara}.


Our results for $W_\text{d}$ and $W_\text{s}$ are presented together
with angular momentum quantum numbers $\Omega$ in \prettyref{tab:
allprops}.  All $\Omega$ values are close to the expected value
$^1/_2$. Minor numeric deviations from $^1/_2$ are due to slight
numeric deviations of the computed equilibrium structure from
linearity and, caused by this, an imperfect alignment on the $z$-axis.

Values calculated for $W_\text{d}$ and $W_\text{s}$ on the DFT level
for group 2 hydroxides differ only slightly from those obtained with
GHF, which is in agreement with previous studies of
$\mathcal{P,T}$-violation in group 2 compounds.\cite{gaul:2018} The
larger deviation for YbOH shows that electron correlation effects may
be more important for this f-block compound. However, previous
comparisons of our method with coupled-cluster four-component
calculations shows that the accuracy of the present approach can be
estimated to be on the order of 20~\%, which is fully sufficient for
the present purpose.

We find that compared to $\mathcal{P,T}$-odd enhancement in metal
fluorides, calculated in Ref. \cite{gaul:2018}, the values for the
corresponding hydroxides are slightly larger in magnitude, but all in
all differences are very small, below 5\%. Considering possible
improvements of the experimental setup with polyatomic molecules as
described in Ref. \cite{kozyryev:2017a}, experiments with
laser-coolable RaOH or YbOH as promising candidates for an improvement
of current limits on the eEDM consequently would benefit mainly from
full polarisation of the molecule and the existence of internal
co-magnetometer states, but not from improved electronic enhancement
factors. The potential of the latter in polyatomic molecules is thus
yet to be explored.

\begin{figure*}\centering
\resizebox{\textwidth}{!}{\includegraphics{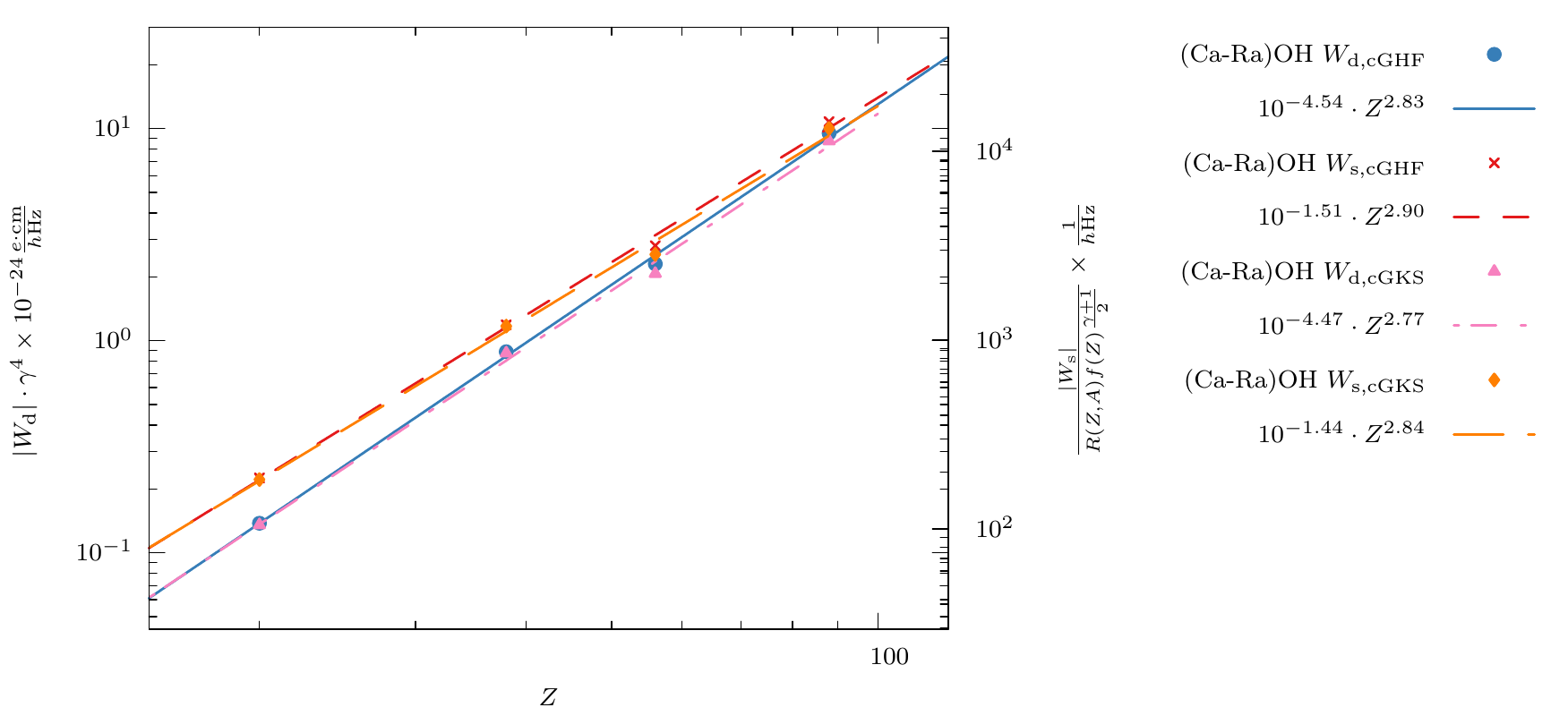}}
\caption{
Scaling of
$\log_{10}\braces{|W_\text{d}|\gamma^4\times10^{-24}~\frac{e\cdot\mathrm{cm}}{h\mathrm{Hz}}}$
and  $\log_{10}\braces{\frac{|W_\text{s}|}{R(Z,A)f(Z)\frac{\gamma+1}{2}}\times\frac{1}{h\mathrm{Hz}}}$ 
with $\log_{10}\braces{Z}$ for group 2 fluorides (Ca-Ra)F at the level of GKS-ZORA/B3LYP and
GHF-ZORA.}
\label{fig: scaling}
\end{figure*}

For further insight the scaling with nuclear charge $Z$ is studied.
For this purpose non-linear relativistic enhancement is separated as
described in Ref. \cite{gaul:2018} using relativistic enhancement
factors known from atomic
considerations\cite{dzuba:2011,fermi:1933a,fermi:1933}
$R_\text{s}=R(Z,A)f(Z) \frac{\gamma+1}{2}$ and
$R_\text{d}=\frac{1}{\gamma^4}$ with
$f(Z)=\frac{1-0.56\alpha^2Z^2}{\parantheses{1-0.283\alpha^2Z^2}^2}$
and
$R(Z,A)=\frac{4}{\Gamma^2\parantheses{2\gamma+1}}\parantheses{2Zr_\text{nuc}/a_0}^{2\gamma-2}$.
Here $\alpha$ is the fine structure constant, $a_0$ is Bohr's radius
and $r_\text{nuc}\approx1.2~\mathrm{fm}\cdot A^{1/3}$.  A double
logarithmic plot for cGHF and cGKS results together with a linear fit
is presented in \prettyref{fig: scaling}. The $Z$-dependence of
$Z^{2.83}$ for $W_\text{s}$ (cGKS) is similar to that reported for
group 2 fluorides in Ref. \cite{gaul:2018} of $Z^{2.79}$ for
$W_\text{s}$ (cGKS), whereas $W_\text{d}$ scales steeper for MOH
($Z^{2.77}$) than for group 2 fluorides (Ref. \cite{gaul:2018}:
$Z^{2.57}$).

As discussed in detail in Ref. \cite{gaul:2018} the ratio
$W_\text{d}/W_\text{s}$ of two different experiments determines if the
experiments are complementary or redundant for a parallel
determination of $k_\text{s}$ and $d_\text{e}$. In \prettyref{tab:
ratios} the ratios $W_\text{d}/W_\text{s}$ are compared to those of
corresponding fluorides determined in Ref. \cite{gaul:2018}.  

The values show that the metal hydroxides fit perfectly in the model
developed in Ref. \cite{gaul:2018}. Hence there is in this sense no
immediate advantage of using a metal hydroxide instead of a fluoride.
With respect to the coverage region in the parameter space of
$k_\text{s}$ and $d_\text{e}$, however, an experiment with MOH would
be able to reduce the size of the coverage region due to the expected
smaller experimental uncertainties. Experiments with the corresponding
MF compounds would become redundant as essentially the same information is
measured.


In this letter we reported the calculation
of enhancements of an electric dipole moment of the electron in simple
polyatomic molecules. Our calculations show that there is no
considerable difference for enhancement factors between fluorides and
hydroxides. This is also true for the ratio $W_\text{d}/W_\text{s}$.
Thus there is no advantage in using MOH alongside MF in experiment as
both experiments yield the same information on the parameter space of
$d_\text{e}$ and $k_\text{s}$. In order to see distinct differences to
diatomic molecules it may be necessary to study more complex and
non-linear polyatomic molecules.  However, together with the
experimental benefits of polyatomic molecules described in Ref.
\cite{kozyryev:2017a} the herein studied molecules are promising
candidates for an improvement of current limits on
$\mathcal{P,T}$-violating effects.

\begin{acknowledgments}
We thank Timur Isaev for inspiring discussions. Computer time
provided by the center for scientific computing (CSC) Frankfurt is
gratefully acknowledged.
\end{acknowledgments}


%
\clearpage

\begin{table}
\begin{threeparttable}
\caption{Molecular structure parameters calculated with in a
quasi-relativistic ZORA approach  at the cGHF and
cGKS/B3LYP level for radical metal-hydroxides MOH
with M = Ca, Sr, Ba, Ra, Yb.}
\label{tab: molpara}
\begin{tabular}{lc
S[table-number-alignment=center,table-format=1.3,round-precision=3,round-mode=places]
S[table-number-alignment=center,table-format=1.3,round-precision=3,round-mode=places]
c
S[table-number-alignment=center,table-format=1.3,round-precision=3,round-mode=places]
S[table-number-alignment=center,table-format=1.3,round-precision=3,round-mode=places]
c
S[table-number-alignment=center,table-format=2.2,round-precision=2,round-mode=places]
S[table-number-alignment=center,table-format=2.2,round-precision=2,round-mode=places]
}
\toprule
M && 
\multicolumn{2}{c}{$r(\text{M}-\text{O})$/\AA}&&
\multicolumn{2}{c}{$r(\text{O}-\text{H})$/\AA}&&
\multicolumn{2}{c}{$\sphericalangle(\text{M}-\text{O}-\text{H})/^\circ$}\\
\cline{3-4}
\cline{6-7}
\cline{9-10}
&&
cGHF&cGKS&&
cGHF&cGKS&&
cGHF&cGKS\\
\midrule
Ca  && 2.00610075909 & 1.97205376512 && 0.932193931023 & 0.953829621336&& 179.9066264034574 & 179.697822591864  \\
Sr  && 2.13443184448 & 2.11017932976 && 0.933205771753 & 0.954518012973&& 179.9862730294334 & 179.9340038270463 \\
Ba  && 2.23905791012 & 2.20734662653 && 0.934855146803 & 0.956492237718&& 179.9324697315709 & 179.9473717655275 \\
Ra  && 2.31549136776 & 2.28860340362 && 0.934647114915 & 0.955985276506&& 179.9319139524936 & 179.9326145033179 \\
Yb  && 2.0829656877  & 2.00193908249 && 0.932664105638 & 0.953200712106&& 179.9186579615903 & 179.9214510693464 \\
\bottomrule
\end{tabular}
\end{threeparttable}
\end{table}

\begin{table*}\centering
\begin{threeparttable}
\caption{Angular momentum and $\mathcal{P,T}$-violating properties
of hydroxide radicals calculated \textit{ab initio} within a
quasi-relativistic two-component ZORA approach at the cGHF and
cGKS/B3LYP level. Dev. refers to the relative deviation between cGHF
and cGKS results.}
\label{tab: allprops}
\begin{tabular}{
l
S[table-number-alignment=center,table-format=3.0]
S[table-number-alignment=center,table-format=3.3,round-precision=3,round-mode=places]
S[table-number-alignment=center,table-format=3.3,round-precision=3,round-mode=places]
c
S[table-number-alignment=center,table-format=3.2,table-figures-exponent=2,round-precision=2,round-mode=places]
S[table-number-alignment=center,table-format=3.2,table-figures-exponent=2,round-precision=2,round-mode=places]
S[table-number-alignment=center,table-format=2.0,round-precision=0,round-mode=places]<{{\si{\percent}}}
c
S[table-number-alignment=center,table-format=3.2,table-figures-exponent=2,round-precision=2,round-mode=places]
S[table-number-alignment=center,table-format=3.2,table-figures-exponent=2,round-precision=2,round-mode=places]
S[table-number-alignment=center,table-format=3.0,round-precision=0,round-mode=places]<{{\si{\percent}}}
}
\toprule
Molecule& {$Z$}& 
\multicolumn{2}{c}{$\Omega$\tnote{*}}&&
\multicolumn{3}{c}{$W_\text{s}\frac{1}{h\cdot\text{Hz}}$}&&
\multicolumn{3}{c}{$W_\text{d}\frac{e\cdot\text{cm}}{10^{24}\cdot h\cdot\text{Hz}}$}\\
\cline{3-4}
\cline{6-8}
\cline{10-12}
&&
cGHF&cGKS&&
cGHF&cGKS&\multicolumn{1}{c}{Dev.}&&
cGHF&cGKS&\multicolumn{1}{c}{Dev.}\\
\midrule
CaOH &  20 & -0.49369 & -0.49906  && -2.1818604672e+02 & -2.1395103833e+02 &  1.94 &&  -1.4379651070e-01&  -1.4111431041e-01&  1.87\\
SrOH &  38 & -0.50000 & -0.50000  && -1.9967125076e+03 & -1.9670566344e+03 &  1.48 &&  -1.0407360373e+00&  -1.0260143979e+00&  1.41\\
BaOH &  56 &  0.48330 &  0.48331  && -8.7919483349e+03 & -7.9086633431e+03 & 10.05 &&  -3.3174991588e+00&  -2.9838012306e+00& 10.06\\
RaOH &  88 &  0.49445 &  0.47109  && -1.5338788950e+05 & -1.4111216430e+05 &  8.00 &&  -2.7548861524e+01&  -2.5320023938e+01&  8.09\\\\                                       
YbOH &  70 & -0.49981 & -0.49506  && -4.1226867631e+04 & -3.0814212982e+04 & 25.26 &&  -1.1395918032e+01&  -8.5356995876e+00& 25.10\\
\bottomrule
\end{tabular}
\begin{tablenotes}
\item[*] The absolute sign of $\Omega$ is arbitrary. However, relative
to the sign of the effective electric field $W_\text{d}\Omega$ it is
always such that $\text{sgn}\parantheses{W_\text{d}}=-1$.
\end{tablenotes}
\end{threeparttable}
\end{table*}

\begin{table}\centering
\begin{threeparttable}
\caption{$\mathcal{P,T}$-odd ratios $W_\text{d}/W_\text{s}$
of hydroxide radicals MOH calculated within a
quasi-relativistic two-component ZORA approach at the 
cGKS/B3LYP level in comparison to ratios of corresponding fluoride 
radicals MF calculated in Ref. \cite{gaul:2018}}
\label{tab: ratios}
\begin{tabular}{
l
c
S[table-number-alignment=center,table-format=1.2,round-precision=2,round-mode=places]
S[table-number-alignment=center,table-format=1.2,round-precision=2,round-mode=places]
}
\toprule
M&&
\multicolumn{2}{c}{$W_\text{d}/W_\text{s}\times10^{-20}~e\cdot\text{cm}$}\\
\cline{3-4}
&&MOH&MF\\
\midrule
Ca &&  6.5956356889628192 & 6.6200153374233132\\
Sr &&  5.215988090820575  & 5.1668988288706605\\
Ba &&  3.7728262048266985 & 3.7825118314723758\\
Ra &&  1.7943190130774572 & 1.7928707642658059\\                                       
Yb &&  2.7700527651269543 & 2.7591277606933186\\
\bottomrule
\end{tabular}
\end{threeparttable}
\end{table}

\end{document}